# The effect of realistic geometries on the susceptibility-weighted MR signal in white matter

Tianyou Xu[a,*], Sean Foxley[a], Michiel Kleinnijenhuis[a], Way Cherng Chen[a,b], Karla L Miller[a]

[a]Oxford Centre for Functional MRI of the Brain (FMRIB), University of Oxford, Oxford, UK
[b]Singapore Bioimaging Consortium, A*STAR, Singapore

*Corresponding author:
   Tianyou Xu
   FMRIB Centre
   John Radcliffe Hospital
   Oxford OX3 9DU
   UK.
   Email: xut@fmrib.ox.ac.uk

**Key words:** R2*, GRE phase signal, magnetic susceptibility modeling, white matter microstructure, myelin

# Abstract

**Purpose:** To investigate the effect of realistic microstructural geometry on the susceptibility-weighted magnetic resonance (MR) signal in white matter (WM), with application to demyelination.

**Methods:** Previous work has modeled susceptibility-weighted signals under the assumption that axons are cylindrical. In this work, we explore the implications of this assumption by considering the effect of more realistic geometries. A three-compartment WM model incorporating relevant properties based on literature was used to predict the MR signal. Myelinated axons were modeled with several cross-sectional geometries of increasing realism: nested circles, warped/elliptical circles and measured axonal geometries from electron micrographs. Signal simulations from the different microstructural geometries were compared to measured signals from a Cuprizone mouse model with varying degrees of demyelination.

**Results:** Results from simulation suggest that axonal geometry affects the MR signal. Predictions with realistic models were significantly different compared to circular models under the same microstructural tissue properties, for simulations with and without diffusion.

**Conclusion:** The geometry of axons affects the MR signal significantly. Literature estimates of myelin susceptibility, which are based on fitting biophysical models to the MR signal, are likely to be biased by the assumed geometry, as will any derived microstructural properties.

# Introduction

Myelin microstructure in white matter (WM) is important for healthy brain function and in neurological disease. In human and animal brains, normal myelin formation supports healthy development and promotes vital processes such as neuroplasticity (1). In contrast, abnormal myelin conditions, such as demyelination, are associated with many forms of neuropathology such as multiple sclerosis (2). Given myelin's important role in brain function, a long-standing goal in human neuroscience has been to noninvasively estimate properties of myelin – its volume fraction in WM, or more specifically, intact myelin volume fraction – from the MR signal.

There are a number of MR-based markers of myelin, including multi-compartment $T_1$, $T_2^*$ and magnetization transfer mapping (3-6). In addition, myelin has a magnetic susceptibility $\chi$ that is offset to its environment. This arises from myelin's unique chemical composition and ordering of phospholipids within the myelin sheath structure. Following empirical works demonstrating that the frequency dependent MR signal (e.g. spectroscopic imaging) may reflect localized differences in magnetic susceptibility $\chi$ (7-9), recent studies have shown that the magnetic susceptibility of myelin strongly influences the gradient echo (GRE) signal, including both signal phase and magnitude (10-14).

Several biophysical models of WM based on myelin microstructure have been used to interpret the measured GRE signal. Factors influencing this signal include relative volume fractions of myelin and intra-/extra-axonal water, g-ratio (thickness of the myelin sheath), magnetization exchange with myelin water, the presence of paramagnetic iron and the magnetic susceptibility of myelin (4,15-18). Moreover, there is recent evidence that myelin exhibits susceptibility anisotropy, where the magnetic susceptibility depends on the orientation of the phospholipids in myelin with respect to the magnetic field, $B_0$ (4,11,14,19-21).

The present work focuses on the specific geometry of the myelinated axon and its effect upon the susceptibility-weighted signal. Existing models use nested cylinders to describe axons, assuming circular geometries (4,14,15). In reality, a diversity of axonal shapes and myelin geometries exist in WM. While simulations using circular shapes benefit from simplicity, the effects of this assumption have not been studied. Given the role that shape has in altering the field perturbations caused by susceptibility-shifted structures, shape is a potential confound in the extraction of microstructure parameters (e.g. myelin thickness).

Myelinated axons perpendicular to the main magnetic field were modeled in two dimensions with several variations. First, we modeled single axons and axon bundles using circular geometries. Next, we modeled the role of myelin shape on the MR signal by distorting circular geometries. We consider more realistic geometries by using a structural template of myelin microstructure derived from electron microscopy (EM) data. Finally, the signal predictions of circular and EM-based geometries are evaluated against data acquired in a mouse model of demyelination.

**Theory**

Previous biophysical models of axons have assumed idealized packings of nested cylinders which are parallel and infinite along one direction and circular in the orthogonal plane (15). These geometric models have been used to create maps of relative magnetic susceptibility that are then used to forward calculate the corresponding local field perturbations (22,23). For isotropic magnetic susceptibility, the field perturbations are generated through point-wise multiplication of a dipole kernel with the susceptibility map in Fourier space, followed by an inverse Fourier transform. This calculation becomes more complicated under magnetic susceptibility anisotropy (4,11,14), which has been suggested to originate from the radial stacking/orientation of the phospholipid bilayers comprising the myelin sheath (Fig. 1a) (24-26). A tensor formulation of the Fourier method is used to incorporate susceptibility anisotropy in the calculation of the microstructural field (14,21).

In this work, the intra-/extra-axonal regions are considered as an implicit reference (zero susceptibility) from which the susceptibility of myelin is offset. The total magnetic susceptibility of myelin can be expressed as the summation of isotropic and anisotropic susceptibility components defined by rank-2 tensors (Figs. 2a-c). Directional susceptibility anisotropy, characteristic of the phospholipid bilayer, is described by a tensor with non-equivalent diagonal components where $\chi_\parallel \neq \chi_\perp$ (Fig. 1a) (14). $\boldsymbol{\chi}$ is transformed into the common reference frame of the axon by a rotation matrix (Fig. 1b). Next, the spatial susceptibility-tensor map is used to forward calculate the corresponding field perturbation using the Fourier expression in Equation 1 (21)

$$\Delta f(r, \phi) = FT^{-1}\left\{\frac{1}{3}\hat{\mathbf{H}}^\mathbf{T} FT\{\chi_\mathbf{R}(r,\phi)\}\hat{\mathbf{H}} - \hat{\mathbf{H}}^\mathbf{T}\mathbf{k}\frac{\mathbf{k}^\mathbf{T} FT\{\chi_\mathbf{R}(r,\phi)\}\hat{\mathbf{H}}}{\mathbf{k}^2}\right\}\bar{\gamma}\mathbf{H} \quad [1]$$

where $\Delta f(r,\phi)$ is the off-resonance frequency in $Hz$, $\chi_R(r,\phi)$ represents the spatial susceptibility tensor map defined in the reference frame of the axon (Fig. 1b), $\mathbf{H}=[\sin(\theta), 0, \cos(\theta)]H_0$ is the applied magnetic field, $\theta$ is the orientation of the fiber to the magnetic field, $\mathbf{k}=[k_x, k_y, k_z]$ is the spatial frequency vector, $\gamma$-bar is the gyromagnetic ratio, and $FT$ is the Fourier Transform. The applied magnetic field $\mathbf{H}$ in Equation 1 is equivalent to $B_0$ of the MRI magnet. Details on the field perturbation calculations are provided in Supporting Materials.

## Methods

**Microstructure model**

We modeled several geometries, including idealized cylinders with circular cross-sections, elliptic cylinders with elliptical cross-sections, and hyper-realistic geometries based on EM.

For a proof-of-principle examination of the role of shape, we performed field perturbations corresponding to a single axon with elliptical cross section and increasing eccentricities, 0, 0.66, 0.80 and 0.87, corresponding to minor-to-major axis ratios of 1, 4/3, 5/3 and 2, respectively. Changes to eccentricity were made without change to myelin or intra-/extra-axonal areas to conserve g-ratio, which was set to 0.7 (27). In circular axons, the g-ratio is the ratio of the inner radius to the outer radius, while for non-circular axons $g = \sqrt{A_i/A_t}$ with intra-axonal and total axonal areas $A_i$ and $A_t$. The effect of rotation on the ellipse with respect to $\mathbf{H}$ is also examined (Figs. 3e and 3f). Simulations were performed on a 500×500 array, spanning 3×3 μm², with $\mathbf{H}$ defined orthogonal to the longitudinal axis of the fiber.

Packed axons were generated on a 4454×4454 array, spanning 37×37 μm² by random close packing of circles (*n*=1434). Circle/axon sizes followed a Gamma distribution (α=5.7) about a mean radius of 0.46 μm to match EM data (*n*=602 axons). The packing algorithm was developed in-house and is described in Supporting Materials. Results of the packing are shown in Supporting Figure S1. The packed axons were also warped to create a second geometry of packed ellipses using an algorithm package, *twirl1.m* (28) (parameters: a = 1.5; b = 0.8; c = 2.0; d = 1/400). This warping transformation conserves the g-ratio and fiber density.

To model susceptibility anisotropy, it is necessary to determine the angle at which phospholipids in the myelin sheath would be oriented with respect to $\mathbf{H}$. This was achieved by segmenting the myelin structure into azimuthally stacked, rectangular quadrilaterals. Further details are described in Supporting Materials (Supporting Fig. S2).

Demyelination was modeled for cylindrical and EM-based geometries by thinning the myelin structure from the inside out, producing a range of g-ratios from 0.70 (normal) to 0.98 (significant loss of myelin). Further details for demyelination simulations are in Supplementary Material.

**Diffusion model**

Diffusion within axonal field perturbations affects the susceptibility-weighted signal for longer echo times (>20ms)(4) and was simulated for both EM and circular geometries. Monte Carlo with 100,000 spins were conducted in two dimensions, given that field perturbations are constant in the third direction (along axons). We chose a step time corresponding to 4 pixels in the model geometry (small relative to the space between axons) and the number of steps corresponding to 55 ms (EM model: step time 0.0001 ms, 550000 steps; circular model: step time 0.0001612 ms, 341191 steps). In the non-myelin compartments, the diffusion coefficient was set to match measurements along axons (D=2$\mu m^2$/ms), representing diffusion in the absence of axonal hinderance (29). Diffusion was assumed to be negligible for the myelin compartment (30). The effect of diffusion on each spin is calculated by summing the phase accrual experienced with each time step. To demonstrate the effect of diffusion on modelling, we also present some results without static magnetization (D=0$\mu m^2$/ms). Details on the diffusion simulations, model validation and on calculating the diffusion-weighted signal is in Supporting Materials (Supporting Fig. S3).

**Signal calculation**

The complex MR signal was computed from the field maps, $\Delta Hz(r,\phi)$. The total signal is a summation over the frequencies in the intra-/extra-axonal and myelin compartments:

$$S(t) = \sum_{n=1}^{3} \rho_n e^{-t/T_{2,n}} \iint e^{-i2\pi t \Delta \text{Hz}(r,\phi)_n} r dr d\phi \qquad [2]$$

where $n$ denotes the three compartments, for which proton density $\rho$ and $T_2$ were based on literature values (Table 1). Microstructure simulations model the signal decay associated with

$1/T_2'$ to arrive at the total signal decay $1/T_2^* = 1/T_2 + 1/T_2'$. Signal is sampled from central regions of the field maps for single axons (transparent regions in Figs. 3a-f) and axon bundles (blue circle in Fig. 5a) to avoid edge artifacts. Further details on artifacts and model validation are described in Supporting Materials (Supporting Figs. S4 and S5).

Signal simulations for a multi-echo GRE acquisition (e.g. free induction decay) were performed at echo times from 3 to 55 ms to match the cuprizone imaging experiment (described below). In static simulations, we calculated the signal 100 times between 0 and 55 ms, or every 0.55 ms. Plots in Figures 5g and 5h were simulated to 100 ms to demonstrate some of the effects at longer TE, such as the beating pattern starting at 55ms which is the result of distinct frequency groups in the circular model (red). In diffusion-weighted signal simulations, the signal was calculated after each diffusion step. The signal was calculated over 550000 intervals for the EM model and 341191 intervals for the circular model. The number of intervals is dictated by the input step-time, explained in the section on Diffusion simulations. Further, there was a wrapping of the signal phase (of $2\pi$ radians) at 55 ms in the red curve. The curve that is plotted shows the unwrapped signal phase. The sudden accrual of phase is a feature of the interference at 55 ms, where the signal magnitude temporarily reaches 0.

Calculations of the field perturbations based on 2D input susceptibility maps with array size 4000×4000 takes 15 seconds in Matlab (2015b) using 8GB RAM. Monte Carlo simulations of diffusion took 3 hours using parallel computation on a multi-node cluster (400 simulations with 250 spins each). Signal accumulation of all spins takes an additional 3 seconds.

**Electron microscopy acquisition**

All experiments were compliant the local regulatory and ethical standards regarding animal research. One healthy wild-type mouse was anesthetized and perfused with normal Ringer's solution (Electron Microscopy Sciences, Hartfield, PA) and 2% formaldehyde (31). A cerebellar WM region with circular cross-sectional areas was selected where axons are most perpendicular to the sectioning plane with orientation indicated using an ink marker. The EM image was acquired at 7.1 nm on a 4000×4000 matrix. Myelinated axons and myelin sheaths were hand-segmented. Axon size, fiber density (assuming uniformity in the third dimension) and g-ratio were calculated. Axon radius is calculated as the square root of the area of the axon divided by $\pi$. The EM image and distribution of axon sizes are shown in Supporting Figure S6.

**Cuprizone experiment: acquisition**

Demyelination was studied using a cuprizone mouse model in which ingestion of cuprizone, a copper chelator, leads to oligodendrocyte death and subsequent reversible demyelineation (32,33). C57Bl/6 mice ($n$=9, 8-weeks old) were kept on fed of 0.2% cuprizone *ad libitum* over variable durations over a 42-day period (Fig. 6a) to induce varying degrees of demyelination. Mice were sacrificed after 42 days and imaged. Table S1 lists an approximate correlation of days on cuprizone diet to g-ratio and volume fraction of myelin in WM.

Cuprizone mice were scanned *ex vivo* on a 7T pre-clinical scanner (Bruker Clinscan, Ettlingen Germany) using 4-channel receive and body transmit coils. Imaging used a multi-echo GRE sequence (TE=3-55ms, 4ms echo spacing, TR=1500ms, flip angle 70°, FOV 10×10 mm, matrix 124×124, slice thickness 0.3 mm, 10 averages). Three axial slices were acquired at 0, 4 and 8 mm rostral to the Bregma.

**Cuprizone experiment: analysis**

A region-of-interest (ROI) of the corpus callosum (CC) was manually defined for each mouse using the magnitude image at the first TE. The raw complex GRE signal includes phase wraps and a large spatially varying background field. The background field correction was based on the phase images from the first 5 TEs. Spatial phase wrapping was removed (34) and background fields were estimated using a 2D 'projection-onto-dipole-fields' (35). From the resulting background field estimates at the first 5 TEs, we calculated a voxel-wise linear fit to the phase across TEs ($\phi(TE)=TE*m+b$) in order to extract the component of the phase due to the background field. This linear fitting provides a correction for the raw, complex data in each voxel over all TEs: $\exp(i(TE*m + b))$.

Recent work has suggested that the mean (non-microscopic) susceptibility difference between WM and gray matter (GM) also drive non-local field contributions (14) . Nonlocal field perturbations due to WM/GM tissue interfaces were calculated with the aid of diffusion tensor imaging data, where the latter was used to account for susceptibility anisotropy based on fiber orientation (14). Calculations of the nonlocal perturbations and details of the DTI acquisition are described in Supporting Materials (Supporting Figure S7).

**Cuprizone experiment: model comparisons**

Simulations assuming only isotropic susceptibility were performed to investigate whether isotropic susceptibility was sufficient in modeling the MR signal. Field perturbations of circular and EM models (Figs. 4a and 4c) assuming $\chi_a = 0$ and $\chi_i = -100$ ppb were calculated using Equation 1. The complex MR signal was calculated to 55 ms using Equation 2 assuming microstructure properties listed in Table 1.

Simulations of the MR signal using the circular model (Fig. 4a) under a different $\chi_a$ was performed. Field perturbations were calculated using Equation 1 assuming $\chi_a$ of -70 ppb (as opposed to -120 ppb) and $\chi_i$ to -60 ppb. Field perturbations under nine different g-ratios, ranging from 0.70 to 0.98, were performed. The complex MR signal corresponding to each simulation was calculated using Equation 2 with parameters listed in Table 1.

# Results

**Single axon simulations**

We first investigate the effect of geometry of a single axon on the local magnetic field perturbation. Six different geometric cases and their field perturbations were generated to examine the effect of axon shape and orientation. Figures 3a-d show axons as ellipses of increasing eccentricities, ranging from circular (eccentricity 0) to more elliptical. Figures 3e and 3f demonstrate the effect of rotation of the highest eccentricity ellipse relative to **H**, ranging from orthogonal to parallel. Differences in signal behavior can be attributed solely to the changes in axon shape because volume fractions of myelin, intra-axonal and extra-axonal space are conserved.

The signal magnitude and phase for each ellipse is plotted in Figures 3g and 3h, demonstrating distinct magnitude and phase profiles with particularly pronounced differences in signal phase. The signal phase shows increasing accumulation in first 30 ms as the axon becomes increasingly elliptical. Rotations of a noncircular geometry can also drive significant signal changes: for example, signal phase corresponding to Figures 3d and 3f are opposite in sign after 55 ms despite having identical shape.

**Simulations at the microstructural scale**

Geometries for more realistic simulations of packed axons at are shown in Figure 4. The simulated noncircular geometry is generated by warping the circular template (Fig. 4a) in a manner that conserves g-ratio and fiber density (Fig. 4b). The EM-based geometry is shown in Fig. 4c.

Field simulations corresponding to geometries in Figures 4a-c are shown in Figures 5a-c. For circular axons, the net frequency histogram (black) exhibits three characteristic peaks, each corresponding to a tissue compartment. The intra- and extra-axonal spaces form sharp peaks at -9.6 Hz (blue) and 0 Hz (green), respectively. The myelin compartment (red) is broader with two distinctive humps at ~0 and 25 Hz. Results from the warped template show an overall smoothing and narrowing of the frequency distribution relative to the circular geometries (Fig. 5e), with the three peaks shifted toward zero. For the EM-based simulations, the total frequency distribution shows no distinguishable peaks, but does retain a strong asymmetric shoulder (Fig. 5f).

The effect of noncircular geometries is also reflected in the simulated MR signal. The distinct peaks observed in Figure 4g produce a beating in the signal magnitude around 55 ms (red, Fig. 5g). This beating is attenuated for the warped case (blue) and is extinguished for the EM-based simulation (black) due to the less distinct frequency peaks. The narrower distributions for the warped and EM geometries generate a more slowly varying signal phase compared to the circular geometry (Fig. 5h). Over the 55 ms duration measured in the cuprizone experiment, the phase for the warped and EM cases have accumulated 0.6 rad (34°) in contrast to the circular case, which accrued 1.5 rad (86°).

**Cuprizone mouse model experiment results**

The experimental design and example images for the cuprizone demyelination experiment are shown in Figure 6. Histological staining confirmed reduced myelin (low stain intensity) in mice with long-duration compared to short-duration diets (Figs. 6f and 6g). Example phase images (TE=23 ms) from a healthy mouse (Fig. 6d) demonstrate markedly stronger contrast between GM and CC compared to a mouse on a 37-day cuprizone diet (Fig. 6e). This is consistent with a reduced myelin volume fraction in the CC, rendering the region less diamagnetic. Table S1 provides an approximate correlation between days on cuprizone diet to g-ratio and volume fraction of myelin in WM.

Averaged signal magnitude and phase from the CC (Fig. 6c) is plotted in Figures 7e and 7f. The phase data have been processed to remove macroscopic field inhomogeneities (see Methods). These plots are color-coded by the cuprizone diet duration. There is a clear trend for faster signal decay and greater phase accumulation in mice undergoing short diet durations (and therefore mostly intact myelin). At TE=55ms, the signal magnitude has attenuated to approximately 0.15-0.35 and the signal phase varies from -0.9-0 rad (-51-0°).

**Signal predictions for demyelination**

Figure 7 also presents forward model predictions for the circular and EM geometries (Figs. 4a and 4c). The myelin sheath is eroded incrementally to simulate nine stages of demyelination wherein the g-ratio ranges from 0.70 (healthy myelination) to 0.98 (severe demyelination). These models were calculated for both static magnetization (no diffusion, Figs. 7a and 7b) and diffusing magnetization (Figs. 7c and 7d). All models predict faster signal decay and phase evolution with higher levels of myelination. The circular model predicts somewhat greater signal decay and significantly greater phase evolution than the EM-based model, and the shape of the signal evolution also differs between geometries. Given that these simulations were otherwise matched, these differences suggest that myelin geometry has a significant influence on both signal magnitude and phase across a range of demyelination stages. Finally, results suggest that diffusion has a significant effect on both GRE signal magnitude and phase predicted by the circular model, but considerably less effect on the EM-based signal model. This effect is further illustrated in Supplementary Material (Fig. S3).

In contrast to results where susceptibility anisotropy is included (Fig. 5), frequency distributions for purely isotropic susceptibility have a positive mean (Figs. 8c and 8d) and generate positive signal phase evolutions (Fig. 8e). Anisotropic susceptibility induces a negative field shift associated with the intra-axonal compartment, resulting in negative phase evolution, consistent with experimental data.

The signal model predictions above use literature values for key parameters such as susceptibility ($\chi_a$=-120 ppb), which have some degree of uncertainty. Changing $\chi_a$ to -70 ppb in the circular model predicts MR signal magnitude and phase evolutions similar to experimental data (Figs. 9a and 9b). This single parameter change shifts the intra-axonal frequency peak from -10 Hz (Fig. 5d) to -6 Hz (Fig. 9c) thereby slowing down the signal phase evolution. Signal simulation across the nine g-ratios shown in Figures 9a and 9b does not include the nonlocal WM/GM corrections used in plots in Figures 7a-d.

# Discussion

This work investigates the role of axon shape on the susceptibility-weighted MR signal. Across geometric models, we match the WM microstructural parameters that are known to influence the MR signal ($T_2$, proton density, $\chi_i$ and $\chi_a$, g-ratio, fiber density), which allows us to attribute differences in the signal predictions to geometry. Nevertheless, other WM parameters that are not included in our simulations can also affect the MR signal, including iron-rich oligodendrocytes, as discussed below.

Single axon simulations demonstrate that varying the eccentricity or in-plane rotation of a myelinated axon with respect to the magnetic field alters the MR signal behavior, implying loss of specificity for biophysical properties such as g-ratio. To probe richer geometries at larger scale, we also considered packed axonal bundles. As with single axons, changes to the myelin shape modulate the underlying frequency distribution and the MR signal, such that a multiplicity of MR signals can be generated from packings sharing the same g-ratio and fiber density.

**Implications of simulations**

The frequency distributions for the simulated geometries exhibit a characteristic set of peaks, which become narrower and less distinct as the simulated geometries become more realistic. These results have important implications for methods aiming to quantify myelin properties from the susceptibility-weighted signal. For example, a logarithmic relationship has been derived between the intra-axonal field shift and the g-ratio for the nested cylinder geometry (14), suggesting a possible *in vivo* measure for g-ratio. For our g-ratio of 0.7, our circular axon geometry predicts an intra-axonal field shift of -9.6 Hz (Fig. 5d) that is consistent with the analytic description. However, simulations from the EM-based model predict that individual compartment distributions are blurred and closer to zero offset, resulting in an aggregation in the frequency distribution and a disappearance of this characteristic intra-axonal peak (Fig. 5f). We have demonstrated that these differences introduce significant alterations to susceptibility-weighted MR signal properties like those used to estimate g-ratio.

Assuming that the EM model is a more accurate reflection of the true underlying microstructure, the implication of these results is two-fold. First, attempts to extract microstructure parameters such as g-ratio from the MR signal would need to incorporate the effect of shape. Second, these results suggest that estimates of myelin susceptibility obtained by fitting circular models of myelin geometry to the MR signal are biased. Under an identical set of parameters (including the same g-ratio and fiber density, therefore myelin content) EM and circular models predict different MR signal. If the EM model is a more accurate representation of white matter microstructure, fitting based on the circular model could underestimate myelin content (Fig. 7b).

Our results therefore highlight a challenge for the use of susceptibility signals for estimating biophysical properties: simple models like circular geometries appear invertible but are unlikely to be accurate, while more realistic geometries like EM-based templates are not a particularly practical approach to biophysical modeling.

**Effect of diffusion**

Simulations that include diffusion predict slower signal magnitude decay and slower signal phase accrual in both circular and EM models compared to static simulations. This is consistent with "motional narrowing", in which spins experience less dephasing as a result of random diffusion (36). Diffusion had a larger influence on the signal predictions for the circular model than the EM model (Figs. 5c and 5d) and reduces the difference in signal phase predictions for the EM and circular models (Figs. 7b and 7d). However, the signal magnitude is predicted to be more different between circular and EM geometries when diffusion is included (Figs. 7a and 7c). Supporting Figure S8 provides an alternative to Figure 7, which compares diffusion and static cases directly.

Incorporating diffusion has the primary effect of reducing the more extreme signal predictions for the circular model, as are seen for the most highly myelinated cases (i.e. the solid lines in Figs. 7a and 7b are overall very different from the other lines of the same color in Figs. 7a-d). Previous work has suggested that incorporation of diffusion into the susceptibility-based white matter signal model can produce more accurate estimates of the experimental data (4). Our signal predictions that include diffusion (taking parameters from literature) are in good agreement with the measured signals from the mouse model of demyelination. However, the significance of this agreement should not be interpreted too strongly in light of the dependence of the signal predictions on parameters like susceptibility with some uncertainty (Fig. 9). Rather, the realism of the EM geometries and diffusion simulations provide evidence that these properties are important to accurate signal prediction.

**Limitations of the current work**

Field simulations shown in Figures 5a-c were performed in 2D assuming that all axons are parallel and infinite in the third dimension. This is an important remaining simplification regarding the structure of axons, which in reality undulate and deform along tracts. Moreover, in EM data of mouse brain WM, microstructures other than axons were observed though not included in the simulation. For example, one such structure, iron-rich oligodendrocytes, occupy significant volume fractions in some areas of the WM and can have significant effects on the susceptibility-weighted signal (37).

In comparing signal simulations of demyelination to the cuprizone mouse measurements, we make the implicit assumption that there is a monotonic relationship between the duration of cuprizone feeding and demyelination (see Table S1). The MR signal measurements do not demonstrate a strict monotonicity with feeding duration, but they do demonstrate the expected overall trend if one groups the mice according to short, intermediate and long duration diets. This could in part reflect differences in the feeding behavior of different animals or differences in the neurobiological response to cuprizone.

In modeling demyelination, we made the assumption that fiber density remains unaffected as the myelin sheath is thinned from the inside out. In reality, the mechanism by which myelin clearance occurs is more complex; different stages of demyelination can be characterized by either myelin debris or clearance (38). After myelin loss, the demyelinated axon may be surrounded by enlarged astrocytic processes coupled with an increase of microglial cells (39). Increases in the population of astrocytes have also been reported, although the volume fraction of extra-axonal space is countered by the decrease in oligodendrocytes (33,40). Such changes in the extra-axonal volume fraction as well as the spatial distribution of myelin throughout the demyelination process could affect the signal behavior.

Changes to axonal properties were minimized by perfusion fixation of the mouse brain. Measurements of axon size ($n$=602, see Fig. S4) followed a Gamma distribution with mean of 0.46 µm, in agreement with previous studies (41).

DTI data was used to compute the nonlocal fields from bulk WM/GM distribution. DTI to date provides the best non-destructive measure of fiber orientation available for whole brains. However, there are known shortcomings of DTI, including inaccuracies in areas with multiple fiber populations. We assume that the principal diffusion direction is aligned to the longitudinal axis of the axon and therefore the $\chi_\perp$ component of the susceptibility tensor shown in Figure 1a. Results from susceptibility tensor imaging (STI) in large fiber bundles are consistent with the principal axis in DTI; however, STI in small fiber bundles, as well as some larger fiber bundles such as the superior regions of corpus callosum has been shown to differ from DTI data in some regions (42). Future simulations of nonlocal WM/GM distortion may benefit from incorporating STI data with DTI data.

**Failure of the isotropic-only susceptibility model**

Recent studies suggest that myelin exhibits anisotropic magnetic susceptibility and that accounting for this property can provide accurate descriptions of the MR signal modeling, particularly signal phase. Our results show that without susceptibility anisotropy, the overall frequency distribution is positive and predicts a positive phase evolution that is not observed in experimental data (Fig. 8). Incorporating susceptibility anisotropy and/or nonlocal bulk WM/GM field perturbations may produce an overall negative frequency distribution, which would in turn predict negative phase evolutions as seen in the data.

**Circular geometries versus EM geometries**

Parameter values used in the simulations were based on literature (Table 1) (14). Under these specific values, results from the EM model provide greater agreement with the measured data. However, there is still considerable uncertainty about the magnetic susceptibility of myelin, which has a significant effect on the signal prediction. Results in Figure 9 suggest that both circular and EM models can produce MR signal behavior that agrees with measurements. Moreover, the plots in Figure 9 did not include nonlocal WM/GM corrections. The only sense in which one can consider the EM-based model to be "better" is that it is based on more realistic microstructural properties taken from EM measurements. This highlights the problematic nature of using simplified biophysical models of MRI signals to estimate MRI-relevant tissue properties that drive those signals (e.g. myelin susceptibility) *and also* to estimate microstructural tissue properties of neurobiological significance (e.g. g-ratio).

# Conclusion

The results from this work suggest that myelin geometry affects the MR signal. Signal predictions using axon models with realistic geometries, with compartmental properties from literature, were significantly different to circular models and were in good agreement with experimental data from a cuprizone-mouse demyelination model. A powerful application of susceptibility-weighted imaging is the potential to estimate tissue properties such as myelin magnetic susceptibility, fiber density and g-ratio (4,6,14,15). This is sought by fitting biophysical models to the measured MR signal. Our results show that these estimates are likely to be biased by assuming simplified, circular geometric models. Elliptical and EM-based models may provide an opportunity to improve the extraction of such tissue parameters. As such, a careful and thorough understanding of the role of shape in the modulation of the MR signal is essential.

**Acknowledgements**

We would like to thank Dr. Errin Johnson and the Dunn School of Pathology EM Facility for the EM images. We would like to thank Dr. Jason Lerch at the University of Toronto for the DTI data.

**Figure/Table Captions, Main Body**

Table 1. Compartmental properties

| Compartment | $T_2$ (ms) | Proton density $\rho$ | $\chi_{isotropic}$ (ppb) | $\chi_{anisotropic}$ (ppb) |
|---|---|---|---|---|
| Intra-axonal | 50 | 1 | 0 | 0 |
| Extra-axonal | 50 | 1 | 0 | 0 |
| Myelin | 15 | 1/2 | -60 | -120 |

**Table 1.** Compartmental Properties. Isotropic and anisotropic magnetic susceptibility values were based on model estimates in (14). A proton density value of 0.5 was based on the known water content of different WM compartments (43). $T_2$ values for intra-axonal, extra-axonal and myelin water were based on (44) which is in fair agreement with estimates from recent works (4,14).

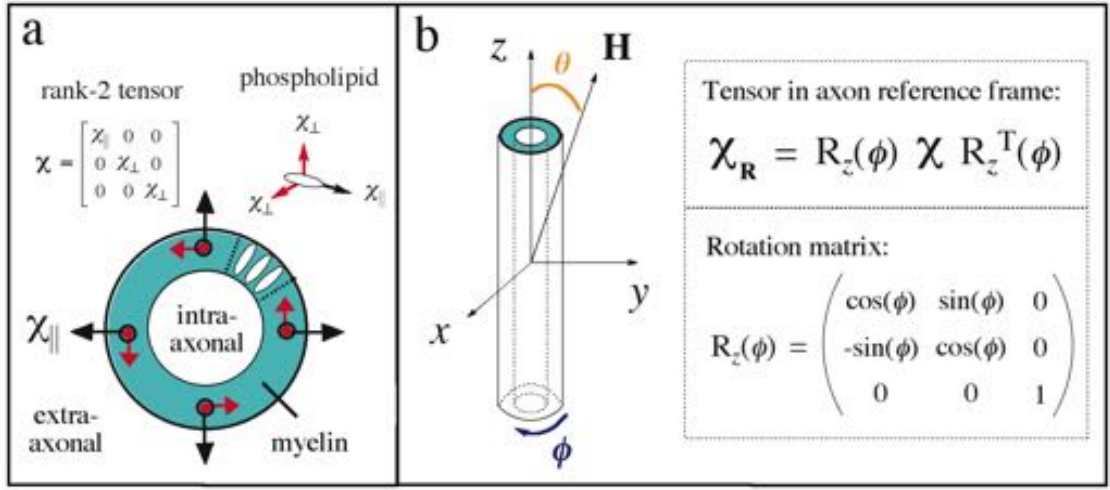

**Figure 1.** (a) The magnetic susceptibility anisotropy of myelin is suggested to originate from its constituent phospholipid bilayer unit, which is a radially oriented in the myelin sheath. Susceptibility anisotropy is described mathematically as a rank-2 tensor. (b) Assuming the longitudinal component of the tensor $\chi_\parallel$ is aligned with $x$, a rotation matrix about $z$ is applied to transform the tensor into the common frame of the axon. Spherical coordinates are used where $\phi$ is the azimuthal angle and $\theta$ is elevation.

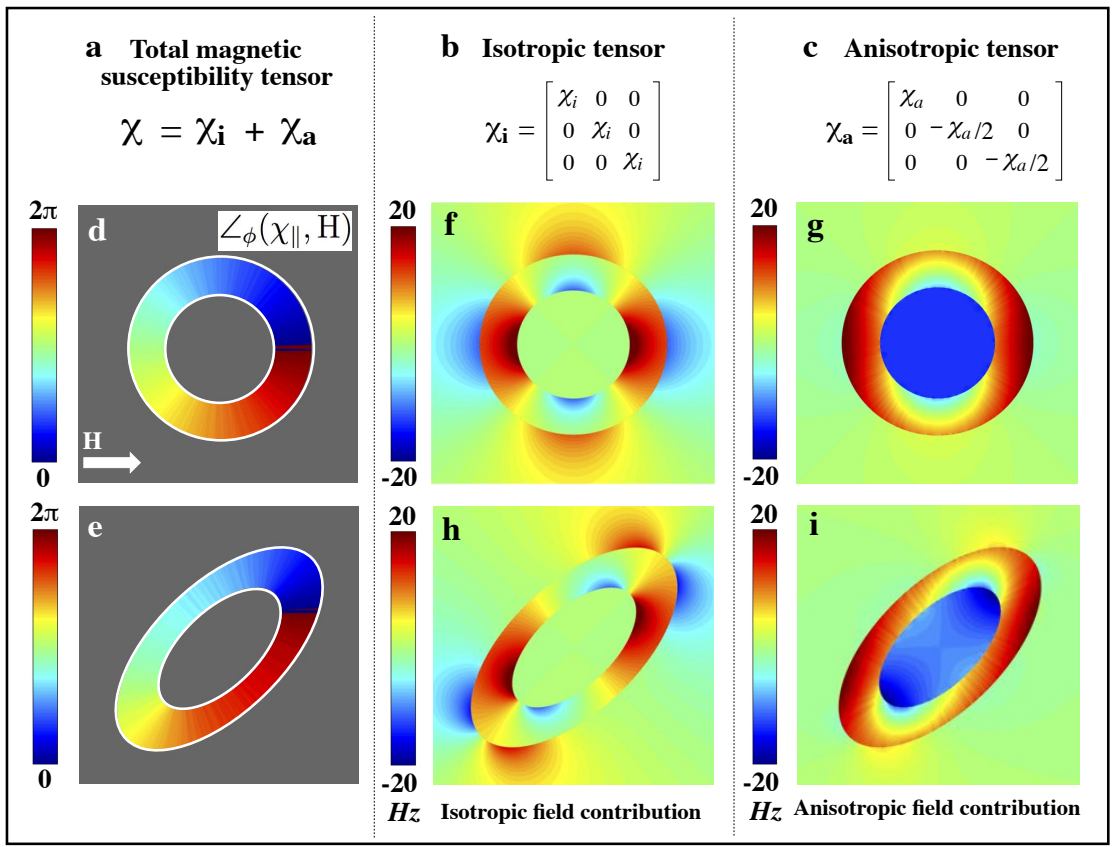

**Figure 2.** (a) The total magnetic susceptibility of myelin can be expressed as the summation of two components: an isotropic component, $\chi_i$ and the anisotropic component, $\chi_a$. The tensor

formulations for $\chi_i$ and $\chi_a$ in the un-rotated frame are shown in (b, c). (d, e) Orientations of the longitudinal tensor component $\chi_\parallel$ with **H** is plotted about the azimuth ($\angle\phi$) for two geometric cases: a nested cylinder and a nested elliptical fiber model. Longitudinal axes of the fibers are assumed to be orthogonal to **H** ($\angle\theta = \pi/2$); perpendicular cross-sections are shown. The isotropic and anisotropic fields corresponding to the nested cylinder are shown in (f, g) and elliptical geometries in (h, i).

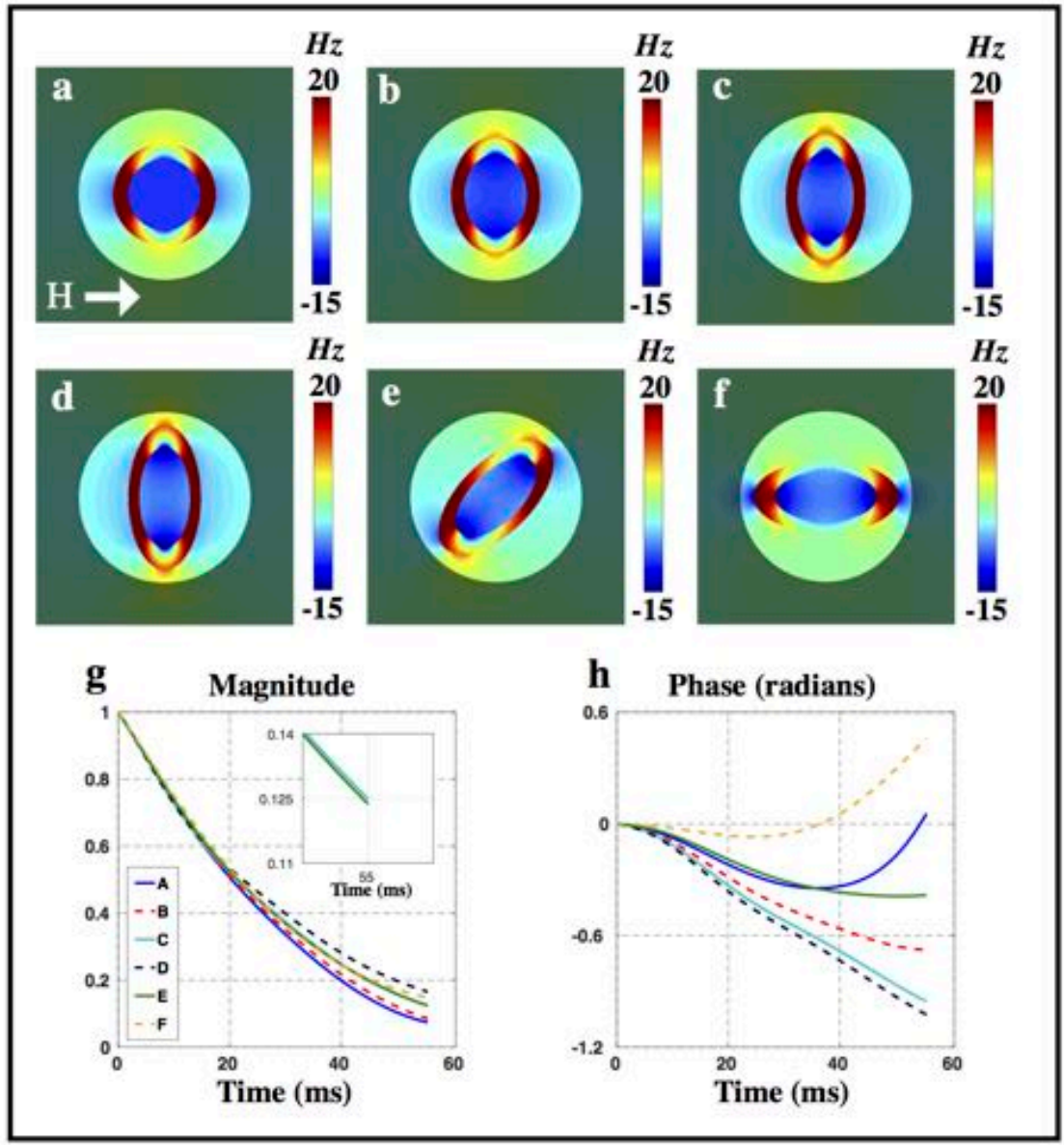

**Figure 3.** (a—d) Field perturbations of elliptical geometries of increasing eccentricity starting from 0 or a circle. (d—f) The effect of in-plane rotations about the applied field. (g, h) The simulated signal magnitude and phase corresponding to cases *a* through *f*.

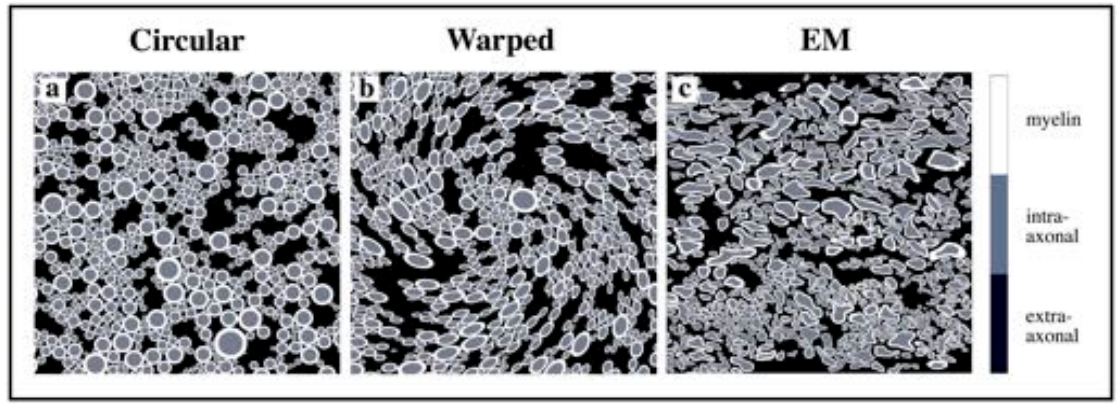

**Figure 4.** Three models of increasing geometric realism are examined: (a) circular axons, *n*=1434, (b) elliptical axons (warped circles), *n*=1434 and (c) axons segmented from EM data of mouse WM, *n*=602. The circular and warped geometries were designed to match relevant properties of the EM segmentation: fiber density and myelin thickness. Myelin structure is shown in white, intra-axonal space in gray and extra-axonal space in black.

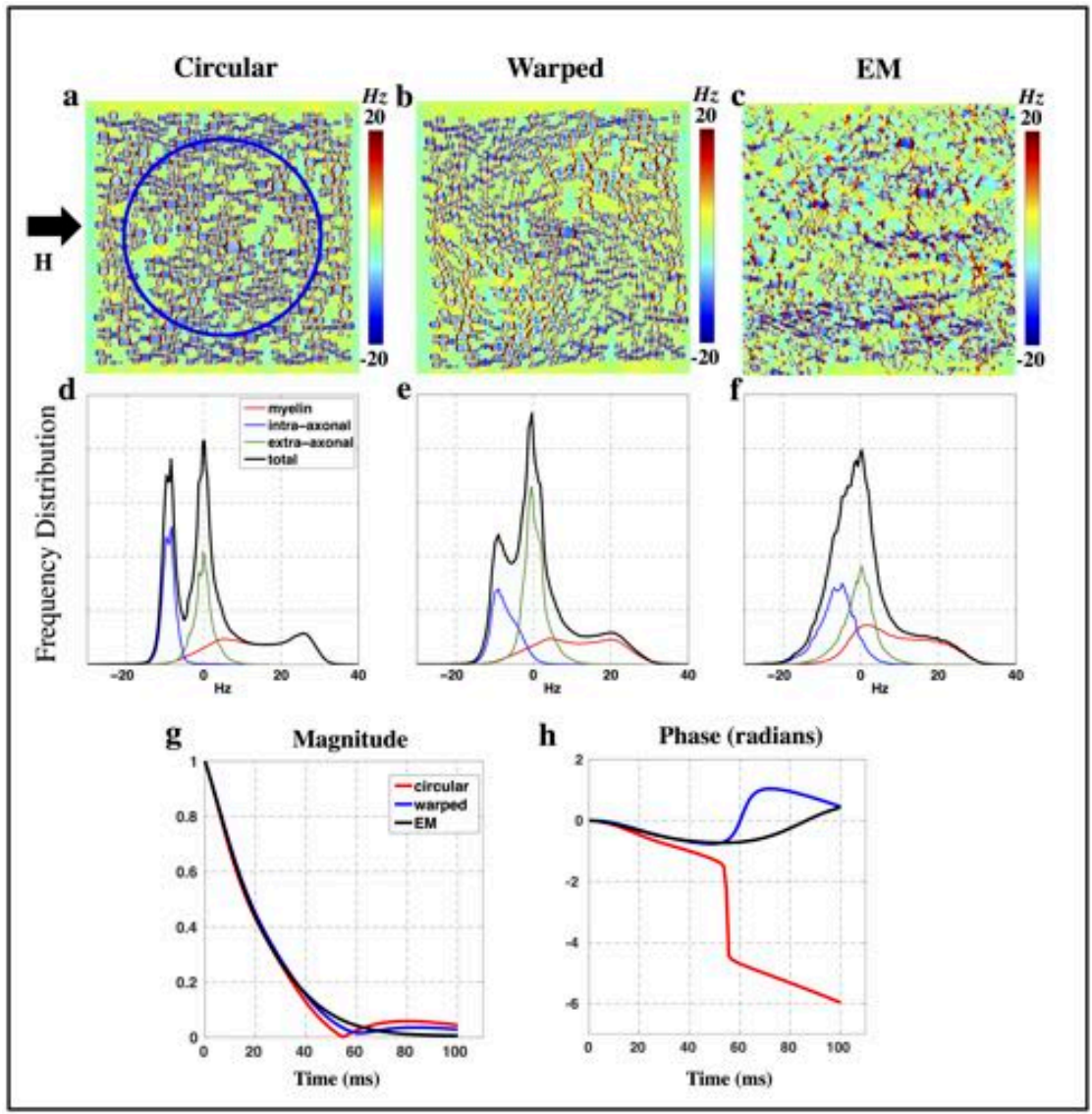

**Figure 5.** Field perturbations corresponding to circular, warped and hyper realistic axon geometries are shown in (a), (b) and (c). Simulations correspond to 7T field strength, with axons orthogonal to the applied field. Axons were assumed to be infinite longitudinally. Corresponding frequency distributions from each simulation are directly below in (d), (e) and (f). Frequency distributions from the intra-axonal, extra-axonal and myelin compartments are shown in blue, green and red respectively. For circular axons, distinct peaks characteristic of the myelin and intra-axonal compartments are visible in the overall distribution. In contrast, the distributions associated with warped and realistic axons are more aggregated with less distinguishable peaks. Comparison of the predicted signal magnitude (g) and phase (h) across the three geometric models: circular axons in red, warped axons in blue, and EM derived axons in black.

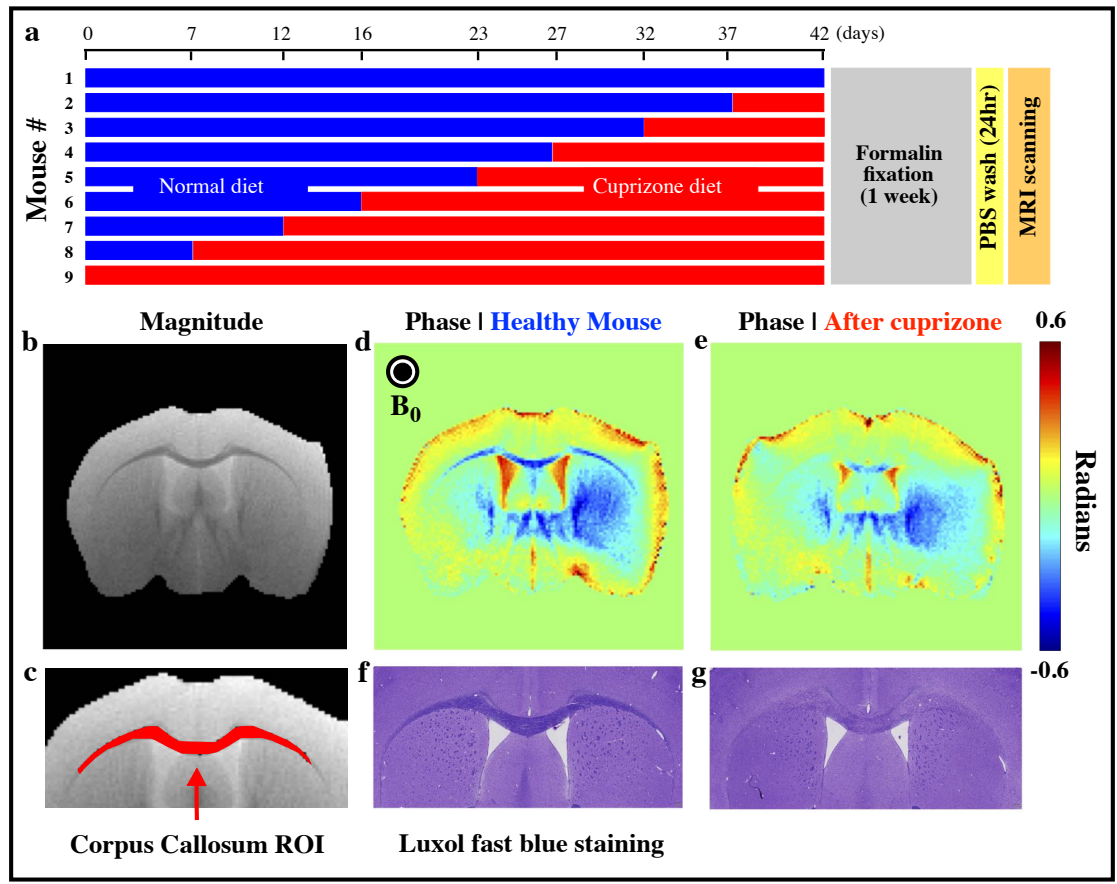

**Figure 6.** Effect of cuprizone on mouse WM. (a) The feeding schedule for nine mice over a 42-day period, followed by sacrifice, fixation and scanning. (b) Axial magnitude image of a mouse with no cuprizone diet. (c) ROI mask over the CC tract, which was used to collect the time-dependent MR signal. (d) Phase image of healthy mouse at TE=23ms. (e) Phase image from the mouse fed a cuprizone diet for 37 days (compared at the same TE=23ms). (f, g) Luxol fast blue histological staining for myelin for the same mice shown in (d,e); healthy myelination marked by high intensity stain in (f) in contrast to lower intensity staining or reduced myelin in (g). Phase wrapping was removed using FSL PRELUDE. Background fields present in the phase images were estimated using the 'projection-onto-dipole-fields' or PDF method in 2D and then removed.

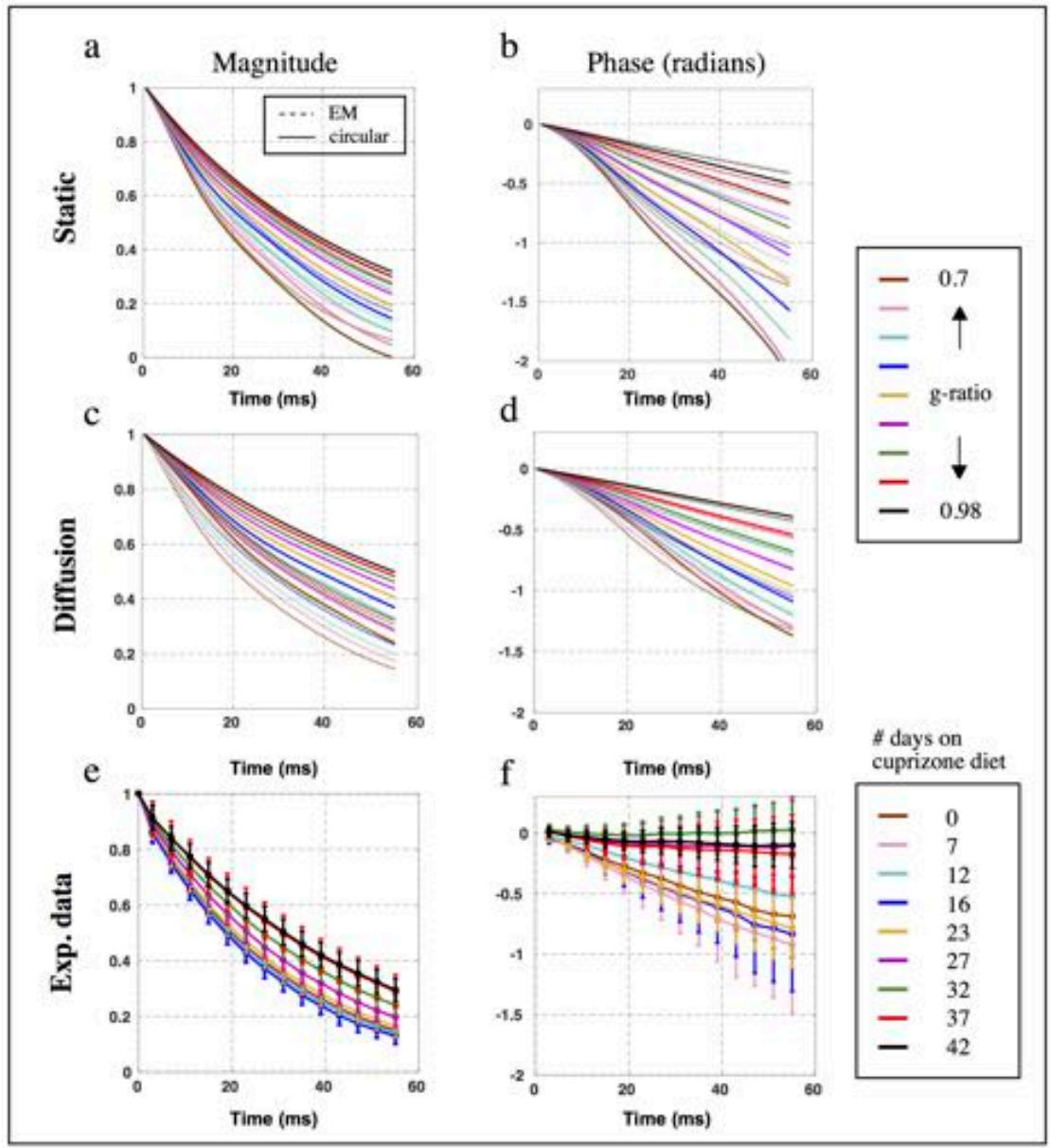

**Figure 7.** Signal modeling of demyelination compared with experimental data. Plots (a-b) and (c-d) show signal magnitude and phase predictions without diffusion and with diffusion. Dotted and solid lines correspond to EM and circular models, respectively. The static magnitude predictions from the circular and EM models (Fig. 7a) range 0-0.32 and 0.07-0.30 at 55 ms, respectively; their static phase predictions range -2.20- -0.50 radians and -1.34- -0.40 radians, respectively (Fig. 7b). The diffusion magnitude predictions from the circular and EM models (Fig. 7c) range 0.23-0.49 and 0.13-0.32 at 55 ms, respectively; their diffusion phase predictions range -1.37- -0.42 radians and -1.40- -0.42 radians, respectively (Fig. 7d). Plots (e) and (f) show the magnitude and phase measured in the cuprizone mouse cohort.

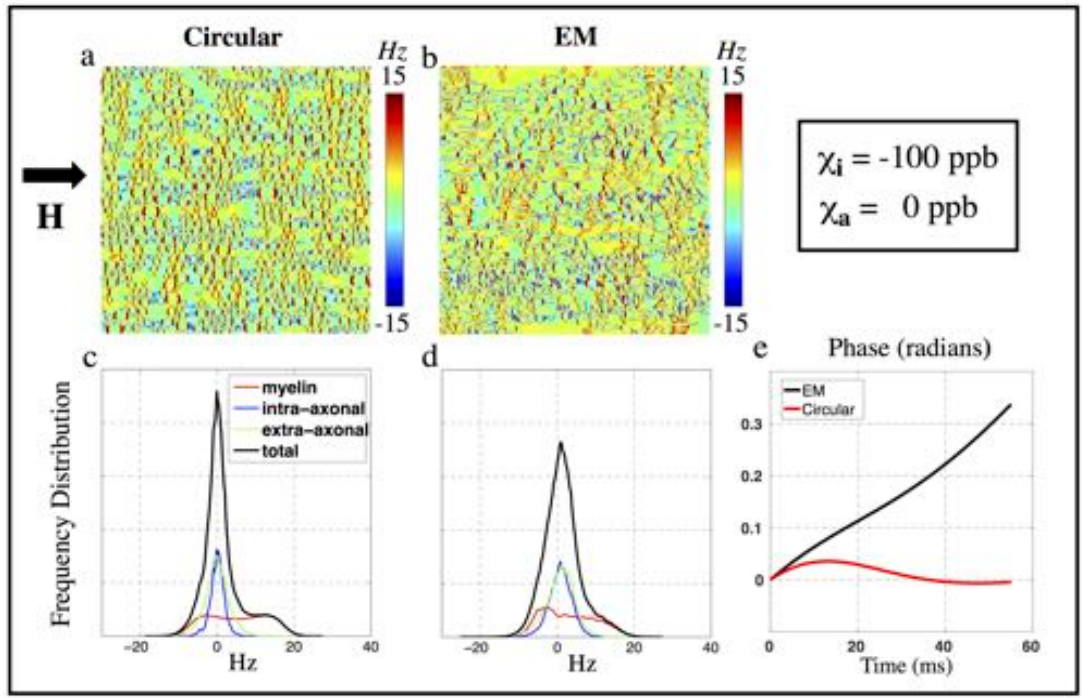

**Figure 8.** Simulations under only isotropic susceptibility, for comparison to the anisotropic model shown in Figure 4. Images (a, b) show the microstructural fields generated from circular and EM geometries. (c, d) show the corresponding frequency distributions for these two examples. The intra-axonal frequency distribution in both models is centered about 0 Hz when only isotropic susceptibility is considered, unlike the anisotropic case. The overall distributions have a positive mean frequency due to the positive shift in the myelin compartment. (e) Simulations predict positive signal phase accrual under purely isotropic susceptibility, unlike the predictions shown in Figure 4.

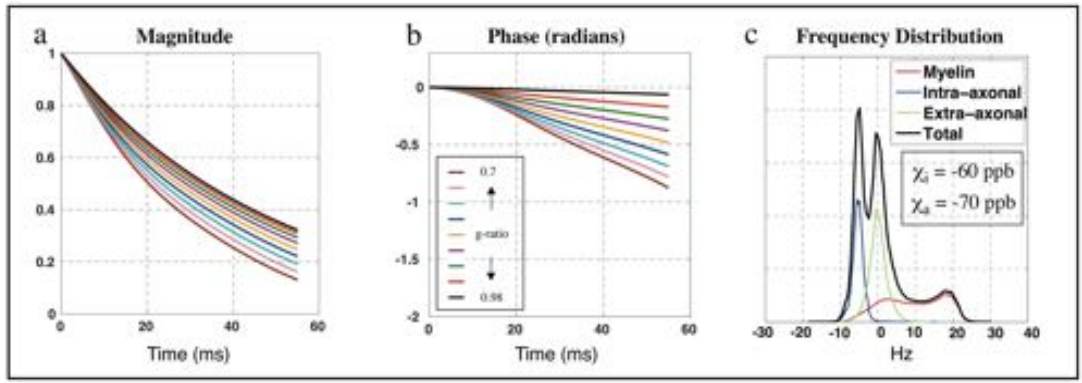

**Figure 9.** Signal predictions for circular geometries for altered susceptibility values can produce similar signal ranges to those observed experimentally. (a, b) The signal magnitude and phase over a range of g-ratios with $\chi_i$=-60 ppb and $\chi_a$=-70 ppb. The result is similar to experimental data and to results from EM models in Figures 7c and 7d which assume $\chi_i$=-60 ppb and $\chi_a$=-120 ppb (c) The change of $\chi_a$ from -120 to -70 ppb shifts the intra-axonal frequency peak from -10 Hz (Fig. 5g) to -6 Hz thereby slowing down the signal magnitude decay and negative phase accrual.

**Supporting Material**

**Theory: field perturbation calculations**

The total field perturbation originating from myelin is the sum of the field contributions from isotropic susceptibility $\chi_i$, where tensor components $\chi_\parallel = \chi_\perp$, and anisotropic susceptibility $\chi_a$, where $\chi_\parallel \neq \chi_\perp$. These tensors, $\chi_i$ and $\chi_a$, in their non-rotated frames are shown in Figures 2b and 2c. The magnitude of $\chi_a$ and $\chi_i$ used in our calculations are based on estimations in previous work (Table 1) (14).

Orientation of $\chi_\parallel$ to the applied field **H** is plotted for two geometric scenarios: orthogonal cross-sections of a "nested cylinder" and "nested elliptical cylinder," shown by the shapes in Figures 2d and 2e. For the isotropic case, where $\chi_\parallel = \chi_\perp$, the susceptibility tensor in the rotated frame (Fig. 1b) becomes an identity matrix. This simplifies the calculation of the frequency field in Equation 1 to the well-known convolution of the susceptibility distribution with a dipole kernel (23). In contrast, field calculations for the anisotropic case rely on explicit information on the orientation of $\chi_\parallel$ to **H**. This is shown by the color-coded maps in Figures 2d and 2e, spanning $0-2\pi$. The field perturbations corresponding to $\chi_i$ and $\chi_a$ are generated in Figures 2f and 2g for the circular geometry and in 2h and 2i for the elliptical geometry. Note that the result in Figure 2f represents the canonical model of cylindrical axons considering only isotropic magnetic susceptibility. In this work, calculations were performed in 2D assuming axon geometries were constant in the third dimension. The orientation of magnetic field, $\theta$, is assumed to be orthogonal to the longitudinal axis of the axon(s).

**Circle packing**

First, circles ($n=1434$) are generated following a Gamma distribution ($\alpha=5.7$) with a mean of 0.46 μm. A circle is selected at random without replacement. The placement or position of the selected circle is determined such that its distance to the average center of all the circles already placed is minimized without overlap. This process is iterated until all circles are placed. The algorithm is uploaded and available here (45). The packing reaches a fiber density of ~83% and is then modified to reach a myelinated fiber density of 64% by random removal of circles and to have an average g-ratio of 0.71. These parameters (density=64%, mean axon size=0.46 μm, g-ratio= 0.71) match the measurements taken from the EM dataset and are also in agreement with literature (27,41). The fiber density measured in the EM data represents only myelinated axons, the population relevant to modeling myelin susceptibility, and is thus lower than would be expected for both myelinated and unmyelinated axons. Results of circle packing is shown in Figure S1.

**Myelin sheath segmentation**

First, the myelin sheath is outlined along its inner circumference and outer circumference. These outlines are then divided ($n=200$) into equally spaced segments. Starting with one point on the inner circumference, a line is made to the nearest point on the outer circumference. Next, a connection is made to the subsequent point (still on the outer circumference). Then, a line is made to a point back on the inner circumference that is one segment away from the original starting point. Finally, the two points on the inner circumference are joined, forming a closed quadrilateral. We assume that the phospholipids traverse the shortest path between the layers of the myelin sheath. The results are shown in Supporting Figure S2. The EM dataset from mouse cerebellar WM is shown in Figure S4.

**Model validation and signal simulation**

Central regions were sampled in single axons simulations for two reasons. The first is due to edge artifacts from discrete Fourier transform operations. A comparison between the analytic solutions, which is assumed to be ground truth, and the Fourier method is shown in the

Supporting Figure S4 highlighting greater inconsistencies near the edges. The second reason is that sampling a central region keeps the volume fraction (assuming continuity in the 3$^{rd}$ dimension) to a more reasonable value. The extra-axonal volume fraction without restricting sampling to the circular FOV would be ~87%. This value is unrealistic for white matter microstructure. Incorporating a FOV reduces the extra-axonal volume fraction to 37% (Figures 3a-f). This is still a high volume fraction according to some literature (46,47). However, this FOV allows for the more eccentric ellipses (Figures 3d-f) to fit within the FOV. The EM model contains fewer axons than the circular bundle model (*n*=602 vs 1434). However, g-ratio and fiber density between the circle and EM models are matched. The FOV is scaled such that total spatial area sampled is consistent across the two models.

We also examined whether the number of axons simulated had an impact on the signal predictions as well whether varying the size of FOV for sampling (and therefore the number of axons) affected signal predictions. Figures S8a and S8b plot the signal magnitude and phase from six separate simulations. The number of axons in these six simulations is color coded, ranging from 1434 to 52. In each simulation a central FOV, which samples 50% of the simulation area, was used to extract the frequencies for signal calculation. Therefore signal predictions come from sampling ~700 to ~25 axons. Results indicate that changes in axon quantity have a minimal impact on signal magnitude and phase compared to a change in axon geometry (Figs. 4g and 4h). In general, simulating a larger number of axons would produce greater accuracy for the model, as it increases homogeneity. Our choice for a model with ~1400 axons was a balance between computation resources, time and accuracy.

Next, we examined whether the size of the sampling FOV affected signal predictions. This was achieved by varying the sizes of the FOV mask on the model of 1434 circular axons and 602 EM axons. Figures S8c and S8d show the magnitude and phase under different FOVs for the circular model (solid lines) and EM model (dashed lines). These lines are color coded to the number of axons sampled within the FOV. The solid black line (C') represents the case where 600 axons are simulated and 300 axons are sampled by the FOV. The results suggest that the size of the central FOV has a negligible effect on the signal magnitude and phase in the circular model. In contrast, we see larger differences in signal phase for the EM case as we vary the FOV size. This is likely because axons in our EM dataset are not as uniformly and homogeneously distributed as is in our circular simulations (Fig. 4c). In the future, it may benefit to use larger EM datasets (>30 μm). There is little variation in signal magnitude for the EM model as FOV is varied. Cross comparison between the circular and EM model results indicate that axonal shape drives the signal changes much more strongly than the number of axons sampled.

**Diffusion simulations**

For the susceptibility model presented here, unmyelinated axons can be neglected when calculating field perturbations under the assumption that their magnetic susceptibility is matched to the extra-axonal space. However, unmyelinated axons exist in abundance and will impede diffusion as their membranes represent boundaries, and thus may have an impact on the calculated signal. To examine whether the inclusion of unmyelinated axons is necessary for model accuracy, we performed additional simulations that packed circular unmyelinated axons with a mean radius of 0.2 μm and standard deviation of 0.05 μm into the extra-axonal space in Figures 4a and 4c to produce Figures S3a and S3b, respectively. Unmyelinated axons have a significantly smaller radius than myelinated axons (8). Diffusion was simulated separately for each of the four compartments (extra-axonal, myelin, intra-axonal, unmyelinated axons) assuming impermeable membranes, and the resulting signal was compared to our standard model without unmyelinated axons. Our results demonstrate that the presence of unmyelinated axons had almost no effect on the signal magnitude or phase (Supporting Figs. S3c and S3d). As such, we use a simpler geometry that does not include unmyelinated axons throughout this paper (Figs. 4a and 4c) for both static and diffusion-weighted simulations.

**Nonlocal field perturbations from WM and GM**

A WM mask was generated by applying a threshold to the fractional anisotropy data (FA>0.25). Next, the principal diffusion direction in each voxel in the WM mask was used to determine the orientation of the principal axis of the susceptibility tensor relative to the applied field, analogous to mapping of the orientation of $\chi_{\parallel}$ to **H** in Figures 2d and 2e. Nonlocal field perturbations were forward calculated in 3D. Experimental measurements from the cuprizone mouse are in 2D, of axial slices through the corpus callosum. We created ROIs of the corpus callosum within the 3D simulation, in axial slices that were anatomically matched as closely as possible to the experimental data. The average field offset $F(Hz)$ was measured from these ROIs and was added to the signal $S(t)$ calculated in Equation 2, which represents the contribution from only the local microstructure. The corrected signal had the form: $S_c(t) = S(t) \exp(i2\pi Ft)$, where $F(Hz)$ was the average field offset. Signal predictions in Figures 7a-d include nonlocal contributions. Calculations of nonlocal WM/GM contributions and microstructure contributions both assume literature values (14,30,44). Their effect on the MR signal is shown in Supporting Figure S7.

Field distortions generated from WM/GM susceptibility contrast is a weighted summation of the isotropic and anisotropic susceptibility field contributions, shown in Equation A1. This equation follows Equation S25 in (14),

$$\Delta \text{Hz} = (v\chi_i + \chi_f)\Delta \text{Hz}_i - \frac{1}{2}v\chi_a \Delta \text{Hz}_a \quad [A1]$$

where $v$ is the volume fraction of myelin in WM, $\chi_i, \chi_f, \chi_a$ are the scalar isotropic susceptibility value of myelin relative to its surroundings, scalar isotropic susceptibility of WM relative to GM and scalar anisotropic susceptibility of myelin, respectively, and $\Delta \text{Hz}_i$ and $\Delta \text{Hz}_a$ are the field perturbations arising from the isotropic susceptibility (Fig. 2b) and anisotropic susceptibility (Fig. 2c), which are calculated using Equation 1. In these nonlocal simulations $\chi_i$=-60 ppb, $\chi_f$=-20 ppb and $\chi_a$=-120 ppb. Volume fraction of myelin in WM or $v$ is calculated as $v = (1-g^2) \cdot d$ where $g$ is the g-ratio and $d$ is density of axons. In our simulations $d = 63\%$. As g-ratio varies from 0.70 to 0.98, $v$ decreases from 0.32 to 0.03, shown in Table S1.

The magnitude of the nonlocal field is a function of $v$. As such, nine simulations were performed with $v$ ranging from 0.32 (healthy) to 0.03 (demyelinated). Simulations were performed in 3D. We created ROIs of the corpus callosum within the 3D simulation in axial slices that were matched anatomically as closely as possible to the experimental data. The ROIs were made using the fractional anisotropy map (FA>0.25). The average field offset is obtained from these ROIs and is shown in Figure S5a, ranging from -1.72 to -0.96 *Hz*. This offset was added to the complex signal, described by Equation 2, arising from microstructure fields (Figs. 4a and 5c) shown in Figure S5b and S5c. For example, the offset from the simulation with $v$=0.32 was added to the signal corresponding to the microstructure simulation where the g-ratio is 0.70. The effect of the nonlocal fields is significant at long echo times. In the EM model, the signal phase from healthy WM (green curve) accrues 0.75 radians in 55 ms without the nonlocal field addition, Figure S5c. The effect of the correction (offset of -1.72*Hz*) causes the signal phase to evolve more rapidly to -1.4 radians in the same time, a nearly 200% change, in Figure S4c.

**Field-of-view in single axon simulation**

Figure S6a shows the field perturbation arising from a single axon (of g-ratio 0.6) generated using the Fourier method described by Equation 1. Figure S6b plots the field perturbations from by their analytic solutions (14). The analytic solutions are assumed to be the ground truth. The difference, by subtraction, between the two fields is shown in Figure S6c, viewed through a colorbar window of -30 to 30 $Hz$. In Figure S6d, this windowing is changed to -2 to 2 $Hz$ to emphasize the edge artifacts which result from the discrete Fourier transforms used in Equation 1. Further, the effect of quadrilateral segmentation of the myelin sheath is accentuated. A circular and central FOV is used avoid sampling the edges of this square array where differences in field are pronounced. Figure S6e and S6f compares the MR signal magnitude and phase arising from the fields in Figure S6a and S6b. The results suggest that the Fourier method offers a fair approximation to the analytic solutions.

**Captions for Supporting Figures**

| # days on cuprizone | g-ratio | volume fraction, $v$ |
|---|---|---|
| 0 | 0.70 | 0.32 |
| 7 | 0.74 | 0.28 |
| 12 | 0.77 | 0.24 |
| 16 | 0.81 | 0.21 |
| 23 | 0.84 | 0.17 |
| 27 | 0.88 | 0.14 |
| 32 | 0.91 | 0.10 |
| 37 | 0.95 | 0.07 |
| 42 | 0.98 | 0.03 |

**Supporting Table S1.** Predicted correlation between days spent on Cuprizone diet, g-ratio and volume fraction $v$ of myelin in white matter.

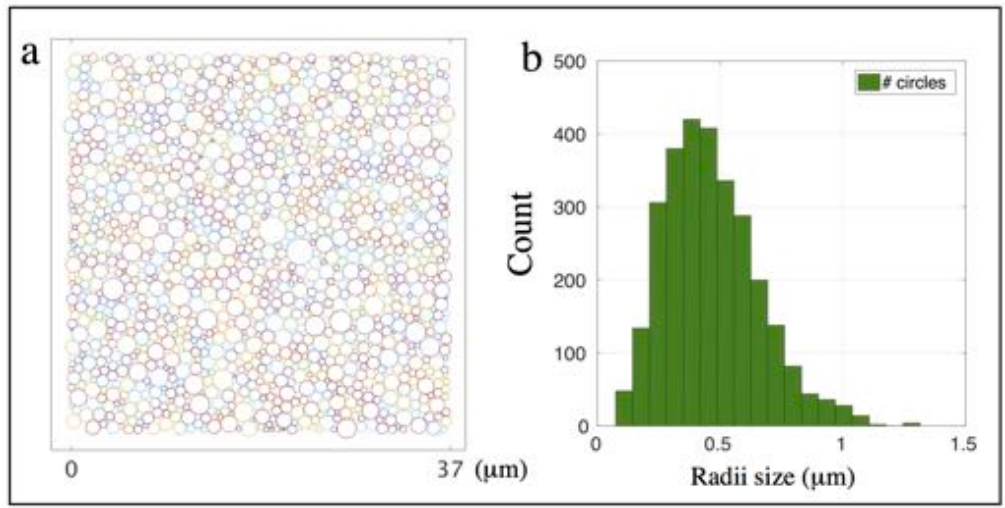

**Supporting Figure S1.** (a) Random close packing of *n*=1434 circles within a square area 37×37 μm². Packing area fraction reaches 83%. (b) Circle radii follow a Gamma distribution with a mean of 0.46 μm, based on literature values.

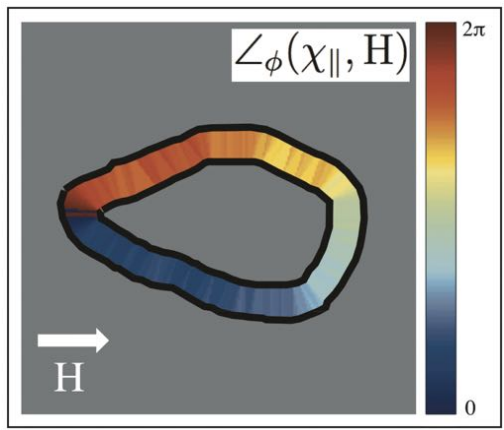

**Supporting Figure S2.** Orientation of myelin phospholipid to the magnetic field in the azimuth plane for a single segmented axon taken from EM data.

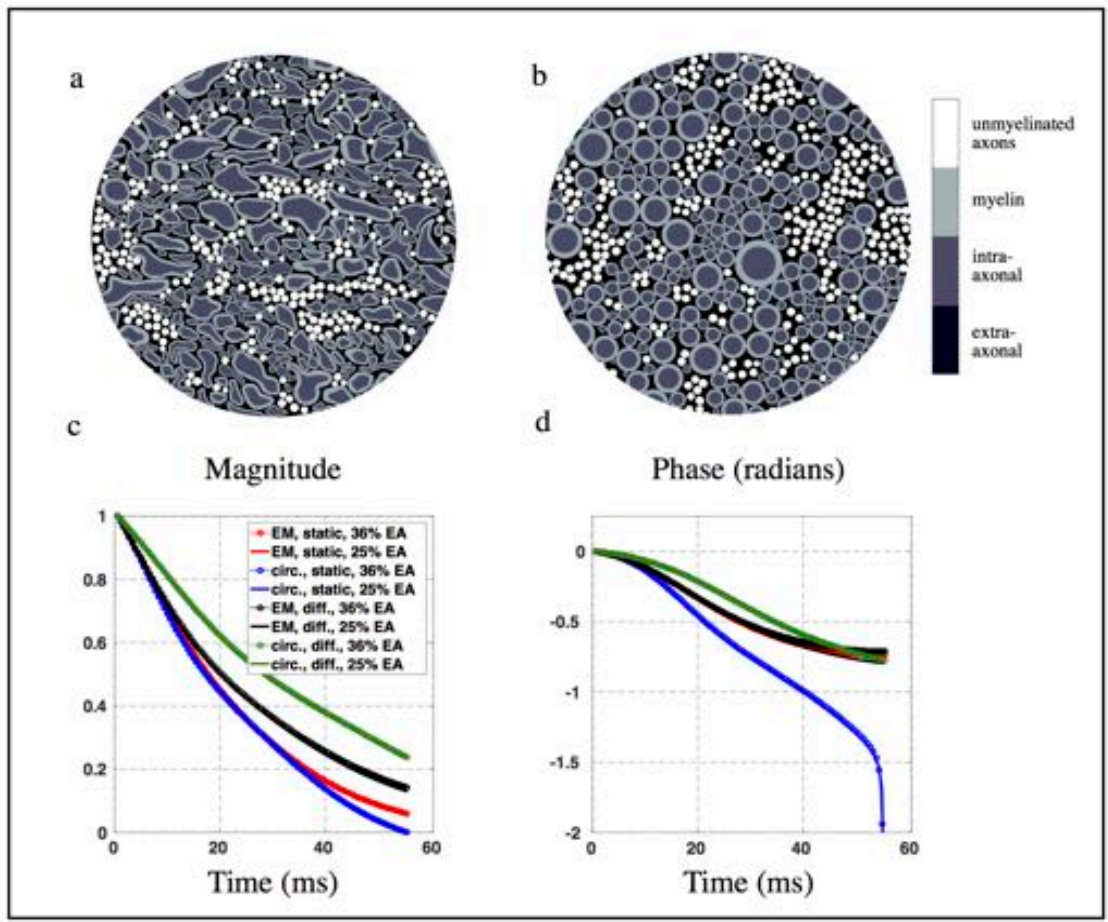

**Supporting Figure S3.** The effect of diffusion is compared for (a) EM and (b) circular models. Unmyelinated axons are packed into the extra-cellular space for more a realistic representation of white matter. These models have an extra-axonal volume fraction 25%, in contrast to 36% associated with the models in Figure 4. The static and diffusion-weighted signal magnitude and phase is plotted in (c) and (d). Results demonstrate that diffusion has a more significant effect on the circular geometry in both signal magnitude and phase. However, unmyelinated axons had little effect on the signal magnitude and phase. As such, we adopt a myelinated-axon model (Fig 4) throughout this work for both static and diffusion weighted simulations.

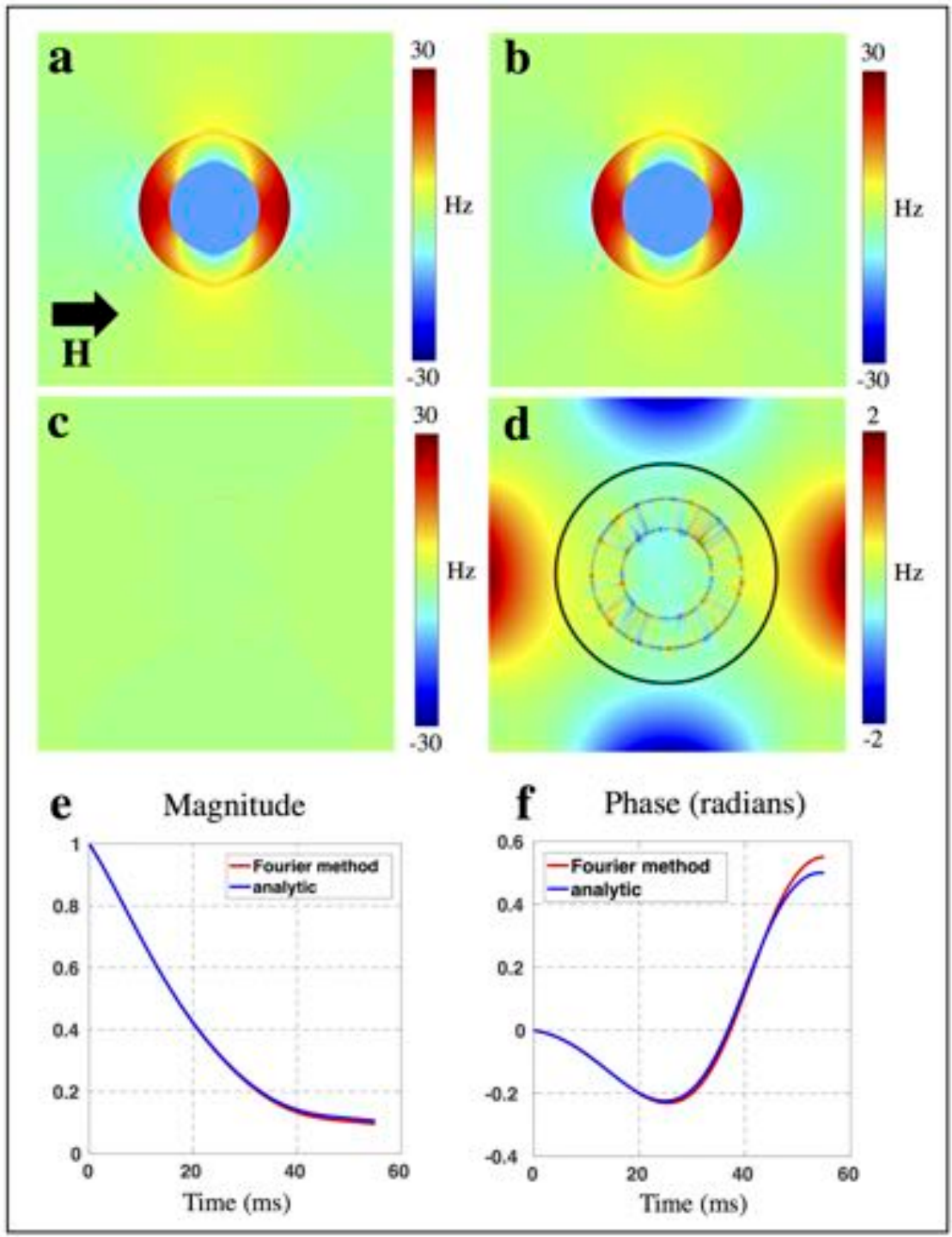

**Supporting Figure S4.** (a) Single axon field perturbation calculated using Fourier method, described by Equation 1, assuming the magnetic field is perpendicular to the longitudinal axis of the axon. (b) Single axon perturbation generated by plotting the analytic solutions or ground truth. (c) Difference between ground truth and Fourier method results at a color bar windowing of -30 to 30 *Hz* (d) Plot of the difference re-windowed to -2 to 2 *Hz* emphasizes edge artifacts from Fourier transform operations and the segmentation of the myelin sheath into quadrilaterals. Outer edge artifacts are avoided by sampling within a central FOV, black

circle. Segmentation-induced artifacts are not avoided. Comparison of the signal magnitude and phase calculated from field perturbations in (a) and (b) with the central FOV is shown in (e) and (f). The results suggest that the segmentation-based Fourier method is a good approximation of the analytic solutions.

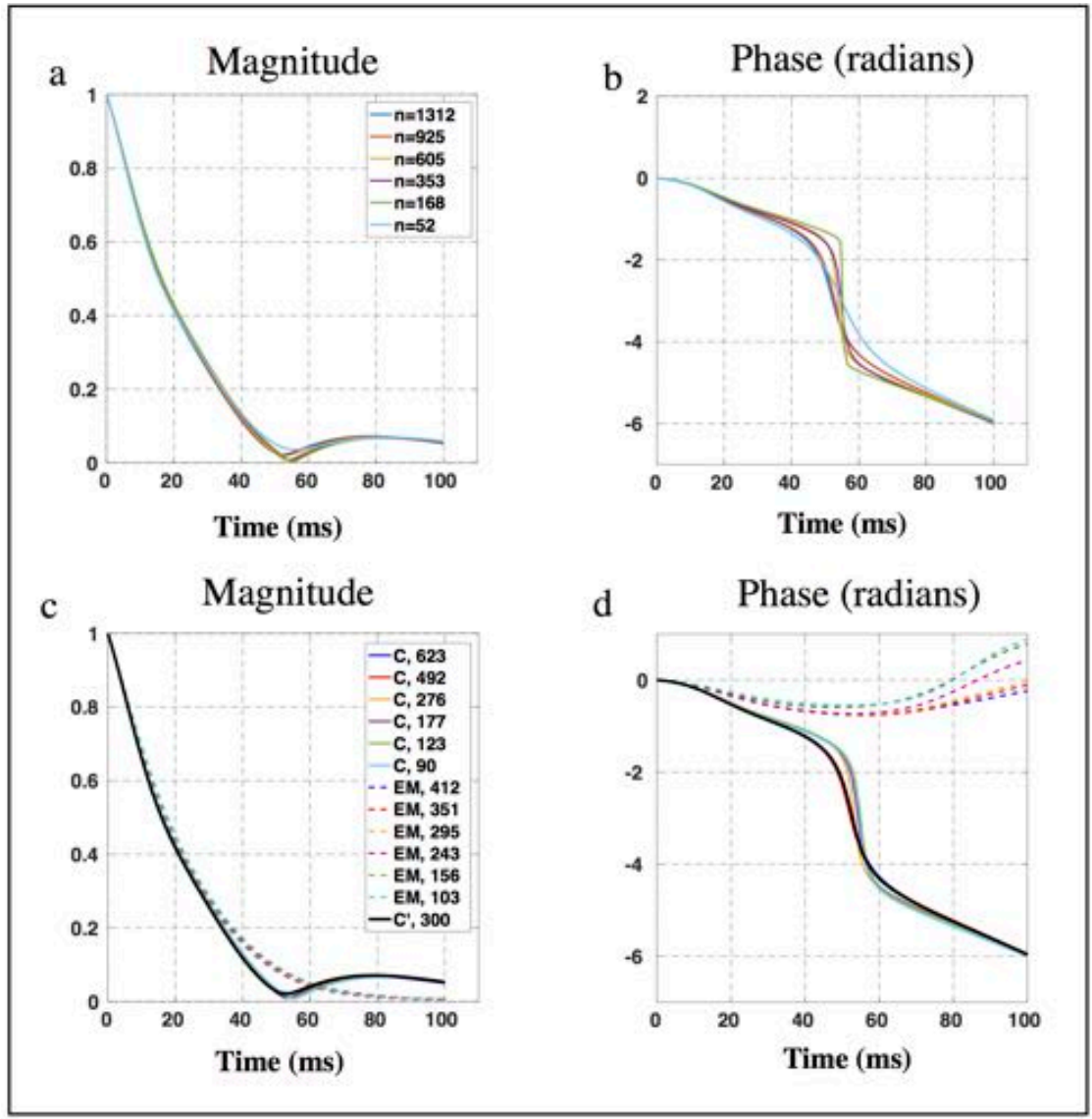

**Supporting Figure S5.** Signal magnitude and phase from six separate simulations are plotted in (a) and (b). The number of axons in these six simulations ranges from 1434 to 52 and is color-coded. In each simulation a central FOV, which samples 50% of the simulation area, was used to extract the frequencies for signal calculation. Panels (c) and (d) compare the signal magnitude and phase between a circular (n=1434) and EM (n=52) model where the size of the FOV is varied. The number of axons sampled within the FOV changes and is color-coded. The black solid line (labeled C') represents the case where 600 circular axons

are simulated and 300 axons are sampled. These simulations suggest that the shape of axons influences the MR signal more than the number of axons simulated as well as the number of axons sampled.

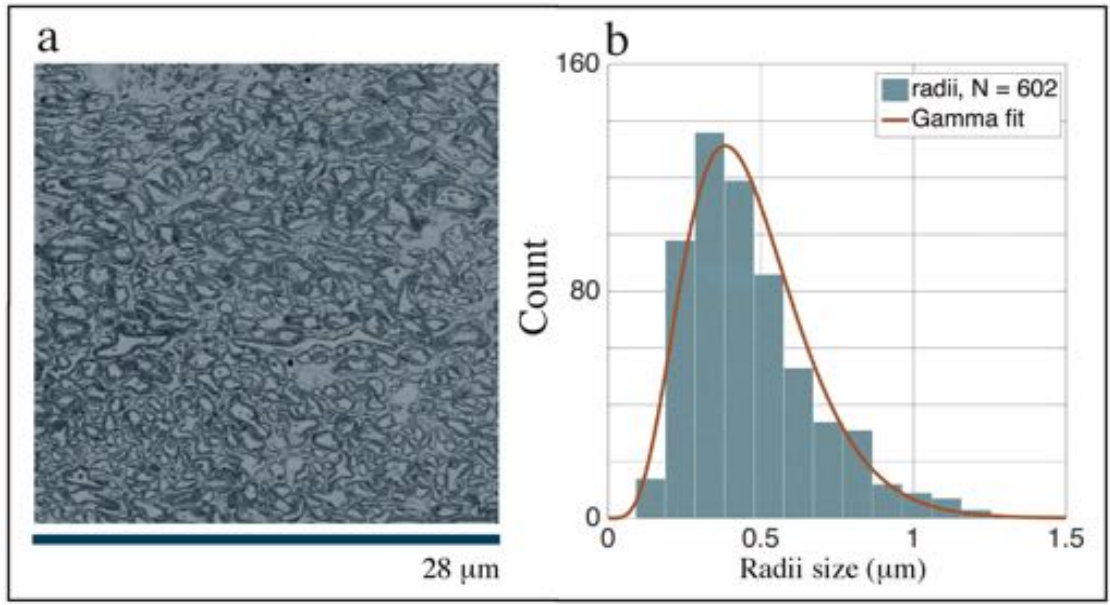

**Supporting Figure S6.** (a) EM image of mouse cerebellar WM, matrix size = 4000×4000 acquired at a resolution of 7.1 nm. (b) Histogram of axon radii size with Gamma fit yielding shape factor, $\alpha$=5.7 and mean radius of 0.46 μm

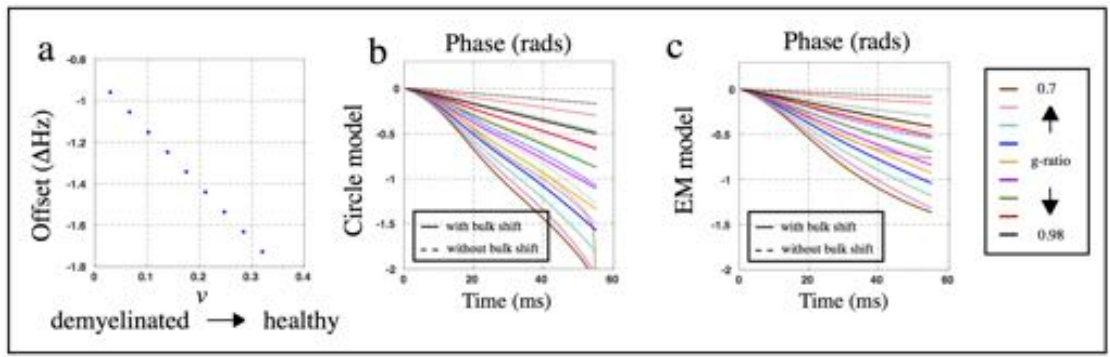

**Supporting Figure S7.** (a) Mean offset in the corpus callosum ROI produced by the nonlocal WM/GM perturbations as a function of myelin volume fraction, $v$, in WM. $v$ ranges from 0.32 (healthy) to 0.03 (demyelinated). (b) Signal phase predictions from circle model with and without nonlocal correction. (c) Signal phase predictions from EM model with and without nonlocal correction.

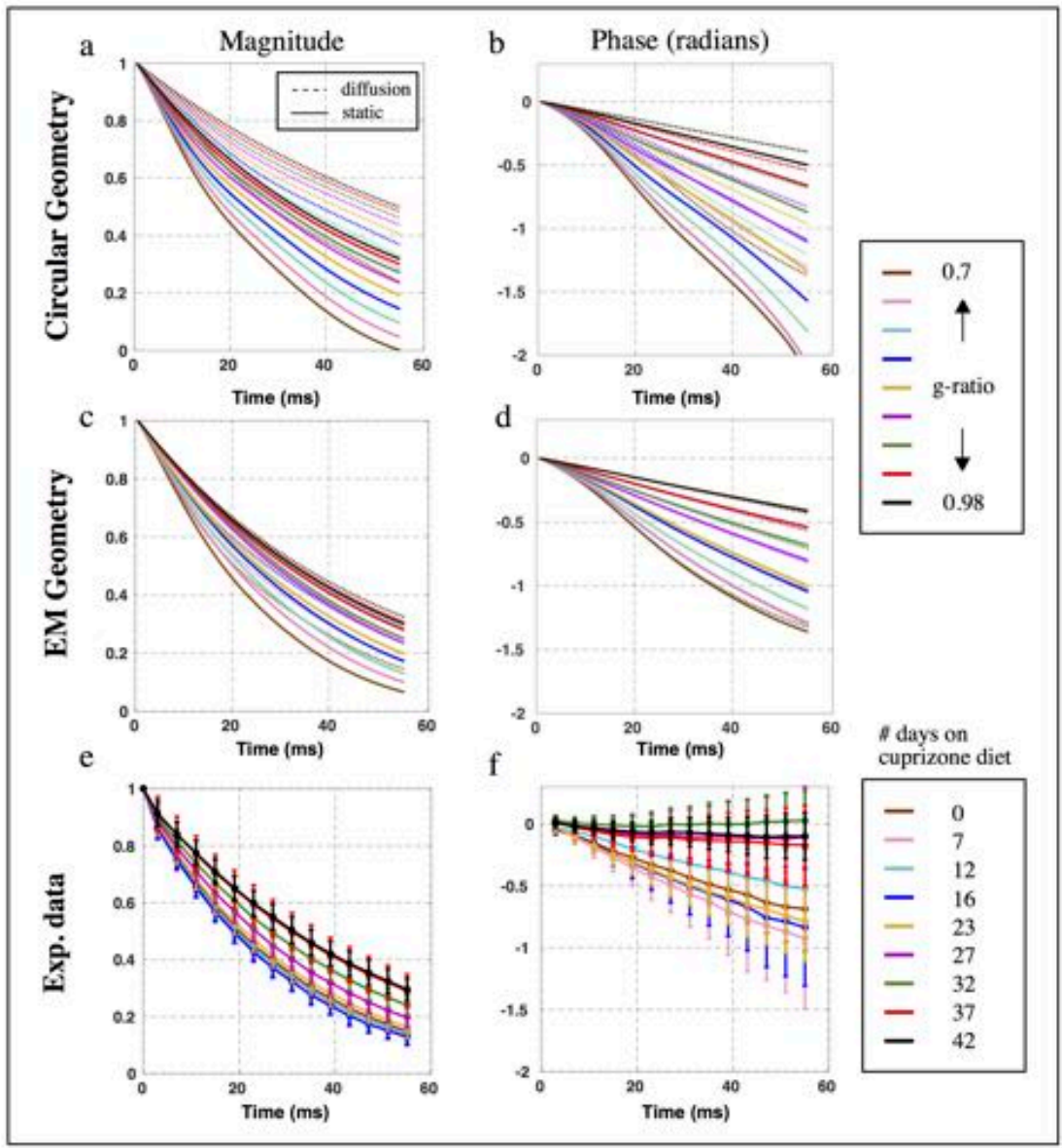

**Supporting Figure S8.** Signal modeling of demyelination compared with experimental data. Plots (a-b) and (c-d) compare signal magnitude and phase predictions across circular and EM models. Dotted and solid lines correspond to diffusion and static results, respectively. Plots (e) and (f) show the magnitude and phase measured in the cuprizone mouse cohort.

**References**

1. Hartline DK. What is myelin? Neuron Glia Biology 2008;4(02):153-163.
2. Steinman, M.D L. Multiple Sclerosis: A Coordinated Immunological Attack against Myelin in the Central Nervous System. Cell;85(3):299-302.
3. Schmierer K, Scaravilli F, Altmann DR, Barker GJ, Miller DH. Magnetization transfer ratio and myelin in postmortem multiple sclerosis brain. Annals of Neurology 2004;56(3):407-415.
4. Sati P, van Gelderen P, Silva AC, Reich DS, Merkle H, de Zwart JA, Duyn JH. Micro-compartment specific $T2*$ relaxation in the brain. NeuroImage 2013;77:10.1016/j.neuroimage.2013.1003.1005.
5. MacKay A, Laule C, Vavasour I, Bjarnason T, Kolind S, Mädler B. Insights into brain microstructure from the $T_2$ distribution. Magnetic Resonance Imaging;24(4):515-525.
6. Laule C, Vavasour IM, Kolind SH, Li DKB, Traboulsee TL, Moore GRW, MacKay AL. Magnetic resonance imaging of myelin. Neurotherapeutics 2007;4(3):460-484.
7. Spees WM, Yablonskiy DA, Oswood MC, Ackerman JJH. Water proton MR properties of human blood at 1.5 Tesla: Magnetic susceptibility, T1, T2, T*2, and non-Lorentzian signal behavior. Magnetic Resonance in Medicine 2001;45(4):533-542.
8. Foxley S, Domowicz M, Karczmar GS, Schwartz N. 3D high spectral and spatial resolution imaging of ex vivo mouse brain. Medical Physics 2015;42(3):1463-1472.
9. Chu KC, Xu Y, Balschi JA, Springer CS. Bulk magnetic susceptibility shifts in nmr studies of compartmentalized samples: use of paramagnetic reagents. Magnetic Resonance in Medicine 1990;13(2):239-262.
10. He X, Yablonskiy DA. Biophysical mechanisms of phase contrast in gradient echo MRI. Proceedings of the National Academy of Sciences 2009;106(32):13558-13563.
11. Lee J, Shmueli K, Fukunaga M, van Gelderen P, Merkle H, Silva AC, Duyn JH. Sensitivity of MRI resonance frequency to the orientation of brain tissue microstructure. Proceedings of the National Academy of Sciences 2010;107(11):5130-5135.
12. Lodygensky GA, Marques JP, Maddage R, Perroud E, Sizonenko SV, Hüppi PS, Gruetter R. In vivo assessment of myelination by phase imaging at high magnetic field. NeuroImage 2012;59(3):1979-1987.
13. Duyn JH, van Gelderen P, Li T-Q, de Zwart JA, Koretsky AP, Fukunaga M. High-field MRI of brain cortical substructure based on signal phase. Proceedings of the National Academy of Sciences of the United States of America 2007;104(28):11796-11801.
14. Wharton S, Bowtell R. Fiber orientation-dependent white matter contrast in gradient echo MRI. Proceedings of the National Academy of Sciences 2012;109(45):18559-18564.
15. Chen WC, Foxley S, Miller KL. Detecting microstructural properties of white matter based on compartmentalization of magnetic susceptibility. Neuroimage 2013;70:1-9.
16. Mackay A, Whittall K, Adler J, Li D, Paty D, Graeb D. In vivo visualization of myelin water in brain by magnetic resonance. Magnetic Resonance in Medicine 1994;31(6):673-677.
17. Fukunaga M, Li T-Q, van Gelderen P, de Zwart JA, Shmueli K, Yao B, Lee J, Maric D, Aronova MA, Zhang G, Leapman RD, Schenck JF, Merkle H, Duyn JH. Layer-specific variation of iron content in cerebral cortex as a source of MRI contrast. Proceedings of the National Academy of Sciences of the United States of America 2010;107(8):3834-3839.
18. Shmueli K, Dodd SJ, Li T-Q, Duyn JH. The Contribution of Chemical Exchange to MRI Frequency Shifts in Brain Tissue. Magnetic resonance in medicine : official journal of the Society of Magnetic Resonance in Medicine / Society of Magnetic Resonance in Medicine 2011;65(1):35-43.
19. Li W, Wu B, Avram AV, Liu C. Magnetic Susceptibility Anisotropy of Human Brain in vivo and its Molecular Underpinnings. Neuroimage 2012;59(3):2088-2097.
20. Sukstanskii AL, Yablonskiy DA. On the role of neuronal magnetic susceptibility and structure symmetry on Gradient Echo MR signal formation. Magnetic resonance in medicine : official journal of the Society of Magnetic Resonance in Medicine / Society of Magnetic Resonance in Medicine 2014;71(1):10.1002/mrm.24629.
21. Liu C. Susceptibility Tensor Imaging. Magnetic resonance in medicine : official journal of the Society of Magnetic Resonance in Medicine / Society of Magnetic Resonance in Medicine 2010;63(6):1471-1477.
22. Salomir R, de Senneville BD, Moonen CTW. A fast calculation method for magnetic field inhomogeneity due to an arbitrary distribution of bulk susceptibility. Concepts in Magnetic Resonance Part B: Magnetic Resonance Engineering 2003;19B(1):26-34.
23. Marques JP, Bowtell R. Application of a Fourier-based method for rapid calculation of field inhomogeneity due to spatial variation of magnetic susceptibility. Concepts in Magnetic Resonance Part B: Magnetic Resonance Engineering 2005;25B(1):65-78.
24. Rosenblatt C, Yager P, Schoen PE. Orientation of lipid tubules by a magnetic field. Biophysical Journal 1987;52(2):295-301.
25. Lonsdale K. Diamagnetic Anisotropy of Organic Molecules. Proceedings of the Royal Society of London Series A, Mathematical and Physical Sciences 1939;171(947):541-568.
26. Lounila J, Ala-Korpela M, Jokisaari J, Savolainen MJ, Kesäniemi YA. Effects of orientational order and particle size on the NMR line positions of lipoproteins. Physical Review Letters 1994;72(25):4049-4052.
27. Chomiak T, Hu B. What Is the Optimal Value of the g-Ratio for Myelinated Fibers in the Rat CNS? *A Theoretical Approach*. PLoS ONE 2009;4(11):e7754.
28. Introduction to Computational Engineering: Morph I2008.
29. P. A. Cook YB, S. Nedjati-Gilani, K. K. Seunarine, M. G. Hall, G. J. Parker, D. C. Alexander. Camino: Open-Source Diffusion-MRI Reconstruction and Processing.  May 2006; Seattle, WA, USA. p p. 2759.
30. Harkins KD, Dula AN, Does MD. Effect of intercompartmental water exchange on the apparent myelin water fraction in multiexponential T2 measurements of rat spinal cord. Magnetic Resonance in Medicine 2012;67(3):793-800.
31. Deerinck TB, Eric;  Thor, Andrea; Ellisman, Mark NCMIR Methods for 3D EM: a new protocol for preparation of biological specimens for serial blockface scanning electron microscopy. 2010.
32. Blakemore WF. Observations on oligodendrocyte degeneration, the resolution of status spongiosus and remyelination in cuprizone intoxication in mice. Journal of Neurocytology 1972;1(4):413-426.
33. Matsushima GK, Morell P. The Neurotoxicant, Cuprizone, as a Model to Study Demyelination and Remyelination in the Central Nervous System. Brain Pathology 2001;11(1):107-116.
34. Jenkinson M. Fast, automated, N-dimensional phase-unwrapping algorithm. Magnetic Resonance in Medicine 2003;49(1):193-197.
35. Liu T, Khalidov I, de Rochefort L, Spincemaille P, Liu J, Tsiouris AJ, Wang Y. A novel background field removal method for MRI using projection onto dipole fields (PDF). NMR in biomedicine 2011;24(9):1129-1136.

36. Bjorkstam JL, Listerud J, Villa M, Massara CI. Motional narrowing of a gaussian NMR line. Journal of Magnetic Resonance (1969) 1985;65(3):383-394.
37. Xu TF, Sean; Miller Karla. Oligodendrocytes and the role of iron in magnetic susceptibility driven frequency shifts in white matter. 2015.
38. Compston A, Coles A. Multiple sclerosis. The Lancet;372(9648):1502-1517.
39. Hiremath MM, Saito Y, Knapp GW, Ting JPY, Suzuki K, Matsushima GK. Microglial/macrophage accumulation during cuprizone-induced demyelination in C57BL/6 mice. Journal of Neuroimmunology;92(1):38-49.
40. Ludwin SK. An autoradiographic study of cellular proliferation in remyelination of the central nervous system. The American Journal of Pathology 1979;95(3):683-696.
41. Barazany D, Basser PJ, Assaf Y. In vivo measurement of axon diameter distribution in the corpus callosum of rat brain. Brain 2009;132(5):1210-1220.
42. Liu C, Li W, Wu B, Jiang Y, Johnson GA. 3D Fiber Tractography with Susceptibility Tensor Imaging. NeuroImage 2012;59(2):1290-1298.
43. Magnetic Resonance of Myelination and Myelin Disorders (2005).
44. Peters AM, Brookes MJ, Hoogenraad FG, Gowland PA, Francis ST, Morris PG, Bowtell R. T2* measurements in human brain at 1.5, 3 and 7 T. Magnetic Resonance Imaging 2007;25(6):748-753.
45. Xu T. Random close packing (RCP) on arbitrary distribution of circle sizes: MATLAB Central File Exchange; 2016.
46. Nicholson C, Syková E. Extracellular space structure revealed by diffusion analysis. Trends in Neurosciences 1998;21(5):207-215.
47. Perge JA, Koch K, Miller R, Sterling P, Balasubramanian V. How the Optic Nerve Allocates Space, Energy Capacity, and Information. The Journal of Neuroscience 2009;29(24):7917.